\documentclass[twocolumn,prl,amsmath,amssymb]{revtex4}
\usepackage{graphicx}% Include figure files
\usepackage{dcolumn}% Align table columns on decimal point
\usepackage{bm}% bold math
\usepackage{latexsym}
\usepackage{amstext}
\usepackage{amsxtra}
\usepackage{color}
\usepackage{epstopdf}
\definecolor{darkblue}{rgb}{0, 0, 0.8}
\usepackage[colorlinks=true, breaklinks=true, linkcolor=darkblue, citecolor=darkblue, urlcolor=darkblue]{hyperref}
\newcommand{\doilink}[2]{\href{http://dx.doi.org/#1}{#2}}

\usepackage{times}

\newcommand{\comment}[1]{}
\begin{document}

\title{An atom-by-atom assembler of defect-free arbitrary 2d atomic arrays}

\author{Daniel Barredo$^*$, Sylvain de L\'es\'eleuc$^*$, Vincent Lienhard, Thierry Lahaye$^\dagger$ and Antoine Browaeys}
\affiliation{Laboratoire Charles Fabry, Institut d'Optique Graduate School, CNRS, Universit\'e Paris-Saclay, 91127 Palaiseau cedex, France\\
$^\dagger$ Corresponding author: \href{mailto:thierry.lahaye@institutoptique.fr}{thierry.lahaye@institutoptique.fr}\\
$^*$ These authors contributed equally to this work}

\date{\today}

\begin{abstract}
Large arrays of individually controlled atoms trapped in optical tweezers are a very promising platform for quantum engineering applications. However, to date, only disordered arrays have been demonstrated, due to the non-deterministic loading of the traps. Here, we demonstrate the preparation of fully loaded, two-dimensional arrays of up to $\sim 50$ microtraps each containing a single atom, and arranged in arbitrary geometries. Starting from initially larger, half-filled matrices of randomly loaded traps, we obtain user-defined target arrays at unit filling.  This is achieved with a real-time control system and a moving optical tweezers that performs a sequence of rapid atom moves depending on the initial distribution of the atoms in the arrays. These results open exciting prospects for quantum engineering with neutral atoms in tunable geometries.
\end{abstract}

\maketitle

The last decade has seen tremendous progress over the control of individual quantum objects \cite{Haroche, Wineland}. Many experimental platforms, from trapped ions \cite{ions} to superconducting qubits \cite{martinis}, are actively explored. The current challenge is now to extend these results towards large assemblies of such objects, while keeping the same degree of control, in view of applications in quantum information processing~\cite{Nielsen}, quantum metrology~\cite{QMet}, or quantum simulation~\cite{QSim}. Neutral atoms offer some advantages over other systems for these tasks. Besides being well isolated from the environment and having tunable interactions, systems of cold atoms hold the promise of being scalable to hundreds of individually controlled qubits. Control of the atomic positions at the single-particle level can been achieved with optical potentials. In a `top-down' approach using optical lattices and quantum gas microscopes, hundreds of traps can now be created and addressed individually \cite{Bakr2008}. By making use of the superfluid to Mott-insulator transition, single atom filling fractions exceeding 90\% are achieved \cite{Zeiher2015}, albeit at the expense of relatively long experimental duty cycles and constraints in the lattice geometries. 

\begin{figure*}[t!]
\centering
\includegraphics[width=18cm]{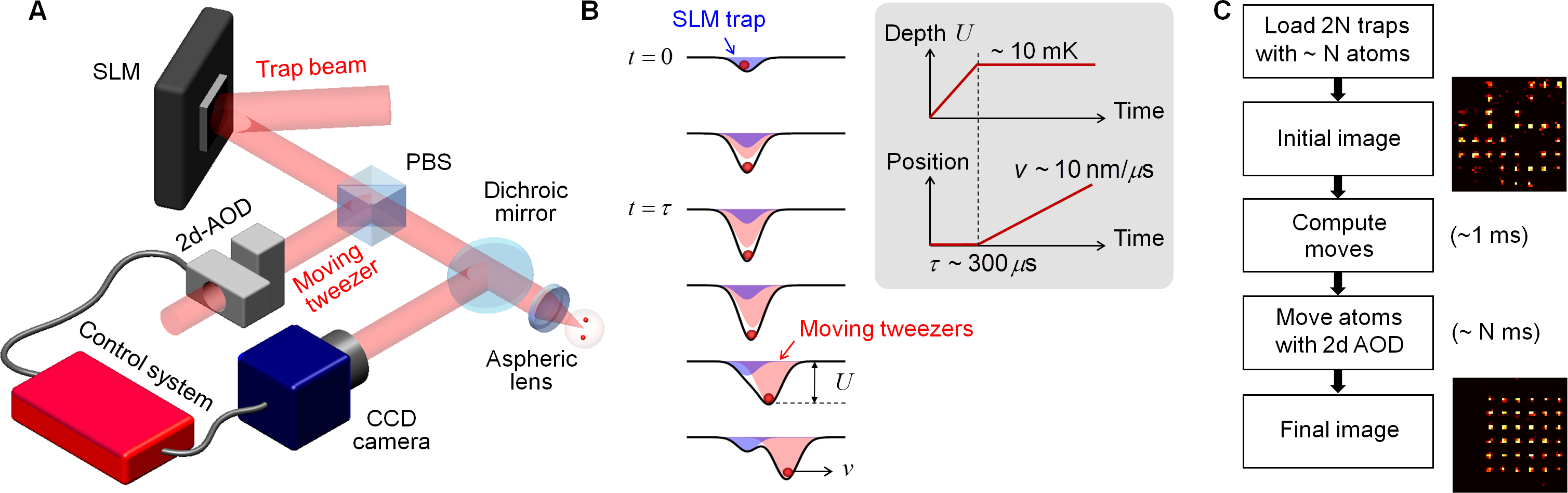}
\caption{{\bf Principle of the atom assembler.} {\bf (A)}: Scheme of the experimental setup. Arbitrary 2d arrays of microtraps are created by imprinting an appropriate phase onto the dipole trap beam using a SLM, and focusing it with a high-NA aspheric lens. The moving beam is generated with a 2d AOD and superimposed on the trap beam with a polarizing beam-splitter (PBS). The CCD camera collects the fluorescence of the atoms in each trap. A control system made of a computer, two microcontrollers, and two AOD drivers allows us to implement a series of rapid moves to reshuffle the atomic array. {\bf (B)}: Extracting a single atom (red disk) from a fixed trap (blue Gaussian) with the moving optical tweezers (red Gaussian). The grey inset shows the time evolution of the moving tweezers depth and position. {\bf (C)}: Block diagram of the control system. Depending on the initial configuration where on average half of the traps are filled, the control system steers each atom towards its target final position. }
\label{fig:fig1}
\end{figure*}

Single atoms can also be trapped in 2d arrays of microscopic optical tweezers with single-site resolution using holographic methods~\cite{Nogrette2014,Schlosser2012,Piotrowicz2013}. This bottom-up approach offers faster preparation and a higher degree of tunability of the underlying geometry. However, achieving unit filling of the arrays is hampered by the stochastic nature of the loading and has remained so far elusive. Although proof-of-principle demonstrations of quantum gates~\cite{Isenhower2010} and quantum simulations~\cite{Labuhn2016} using this latter platform have been reported~\cite{Browaeys2016}, this non-deterministic loading poses a serious limitation for applications where large-scale ordered arrays are required. To solve this problem, several approaches have been considered, exploiting the Rydberg blockade mechanism~\cite{Ebert2014}, or using tailored light-assisted collisions~\cite{Grunzweig2010}. To date, despite those efforts, loading efficiencies of around 90\% at best for a single atom in a single tweezers could be achieved \cite{Fung2015,Lester2015}, making the probabilities to fully load  large arrays still exponentially small. 

A different approach towards this goal, pioneered in Ref.~\cite{Miroshnychenko2006} for a few atoms, and revisited recently in \cite{Lee2016,Kim2016}, consists in sorting disordered arrays of atoms using moving optical potentials~\cite{Beugnon2007}. Here we demonstrate the deterministic preparation of arrays as large as $N\sim 50$ individual atoms in arbitrary 2d geometries, with filling fractions $\eta$ up to 98\%; thus enabling us to achieve defect-free arrays with a fast repetition rate. This is accomplished through the sequential assembly of the atoms in the arrays using a fast programmable control system. Starting from stochastically loaded half-filled arrays with  $\sim 2N$ traps, we analyze in real-time the initial atom distribution, and use a fast, moving optical tweezer to rearrange the atoms into a user-defined target spatial configuration, thus implementing a modern variant of Maxwell's demon~\cite{Weiss2004}. We anticipate that it could be scaled up to a few hundreds of atoms, still maintaining filling fractions close to unity, as the size of the arrays we can prepare is at present mostly limited by the available laser power. These results demonstrate a novel way to initialize arrays of single atoms for quantum simulation \cite{Labuhn2016}.

A scheme of our experimental setup~\cite{Beguin2013,Labuhn2014} is shown in Fig.~\ref{fig:fig1}(A). We use a Spatial Light Modulator (SLM) to create arbitrary two-dimensional arrays of up to 100 traps, separated by distances $a>3\;\mu{\rm m}$ in the focal plane of a high-numerical aperture (${\rm NA}=0.5$) aspheric lens. Each trap has a $1/e^2$ radius of $\sim 1\, \mu$m and a depth of $U_0/k_B \simeq 1\,{\rm mK}$ (for a power of about 5~mW), yielding radial (longitudinal) trapping frequencies around 100~kHz (20~kHz). In the single-atom regime, the traps are stochastically loaded from a magneto-optical trap with cold  single $^{87}{\rm Rb}$ atoms with a probability $p\sim 0.5$. We monitor the occupancy of the traps observing the fluorescence of the atoms at 780 nm with a CCD camera, with a time resolution of 50~ms. For deterministic atom transport we superimpose a second 850-nm laser beam on the trapping beam. We create this moving optical tweezers (with $1/e^2$ radius $ \sim 1.3\; \mu$m) using a 2d acousto-optic deflector (AOD).

Figure~\ref{fig:fig1}(B) shows how we extract an atom from a filled trap using the moving optical tweezer. We first set the horizontal and vertical AOD frequencies to position the beam at the source trap, and ramp up linearly the optical power diffracted by the AODs to reach a tweezers depth $U\sim 10U_0$ in a time $\tau$ (inset). The applied optical potential effectively captures the atom from the trap. Then, we steer the beam towards the target trap, at velocity $v$, by sweeping the vertical and horizontal AOD frequencies. Finally, we release the atom from the tweezers in the target trap. For $\tau\sim0.3$~ms and $v\sim 10\;\mbox{nm}/\mu \mbox{s}$, the probability to succeed in transferring the atoms from the source to the target traps reaches 99.3\% \cite{SOM}. Our method to synthesize fully loaded arrays of $N$ atoms is sketched in Fig.\ref{fig:fig1}(C) and works as follows. We use an array of $\sim 2N$ traps which contains the target array as a subset, load it from the MOT, and trigger the sequence as soon as at least $N$ traps are filled with single atoms. Then, the loading of the array is stopped, and a fluorescence image is acquired to record the initial position of the atoms. Following the analysis of the image, an algorithm (see below) computes on the fly a list of individual atom moves which can rearrange the configuration into the desired pattern. This list is then sent to micro-controllers via serial port communication. The micro-controller program converts this list into a series of voltage sweeps to control the RF drivers driving the AODs. Finally, after the rearrangement operation is completed (in about 50~ms for $N\sim 50$) a final image is acquired to reveal the new positions of the atoms in the array.

\begin{figure*}
\centering
\includegraphics[width=14.5cm]{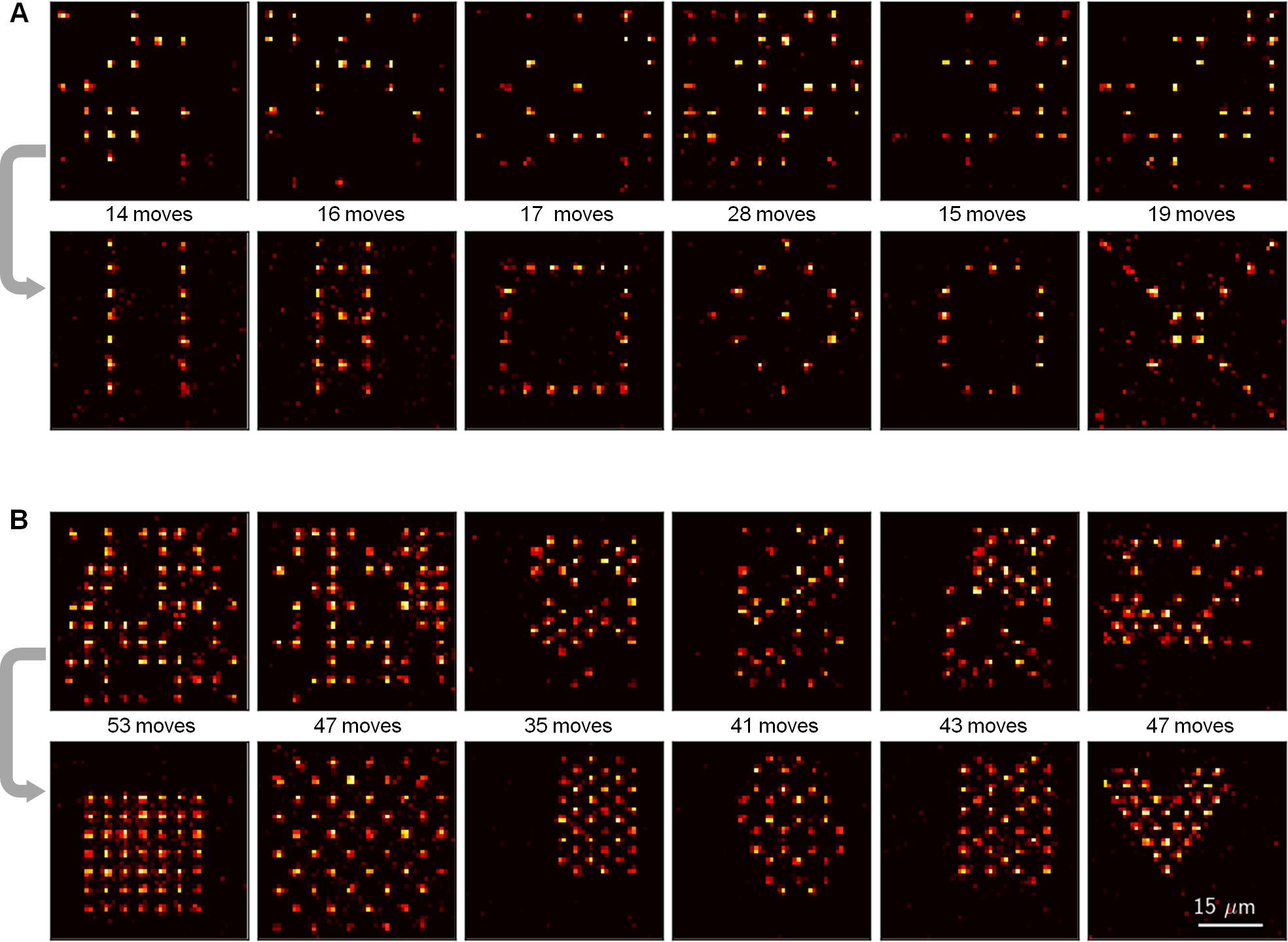}
\caption{{\bf Gallery of fully loaded arrays with arbitrary, user-defined geometries (bottom images) obtained from the initial, random configurations (top images).} All images are single shots. {\bf (A)}: ``Type-1'' moves were used; {\bf (B)}: ``Type-2'' moves were used. The number of elementary moves needed to achieve the sorting are indicated. }
\label{fig:fig2}
\end{figure*}

\begin{figure}
\centering
\includegraphics[width=6cm]{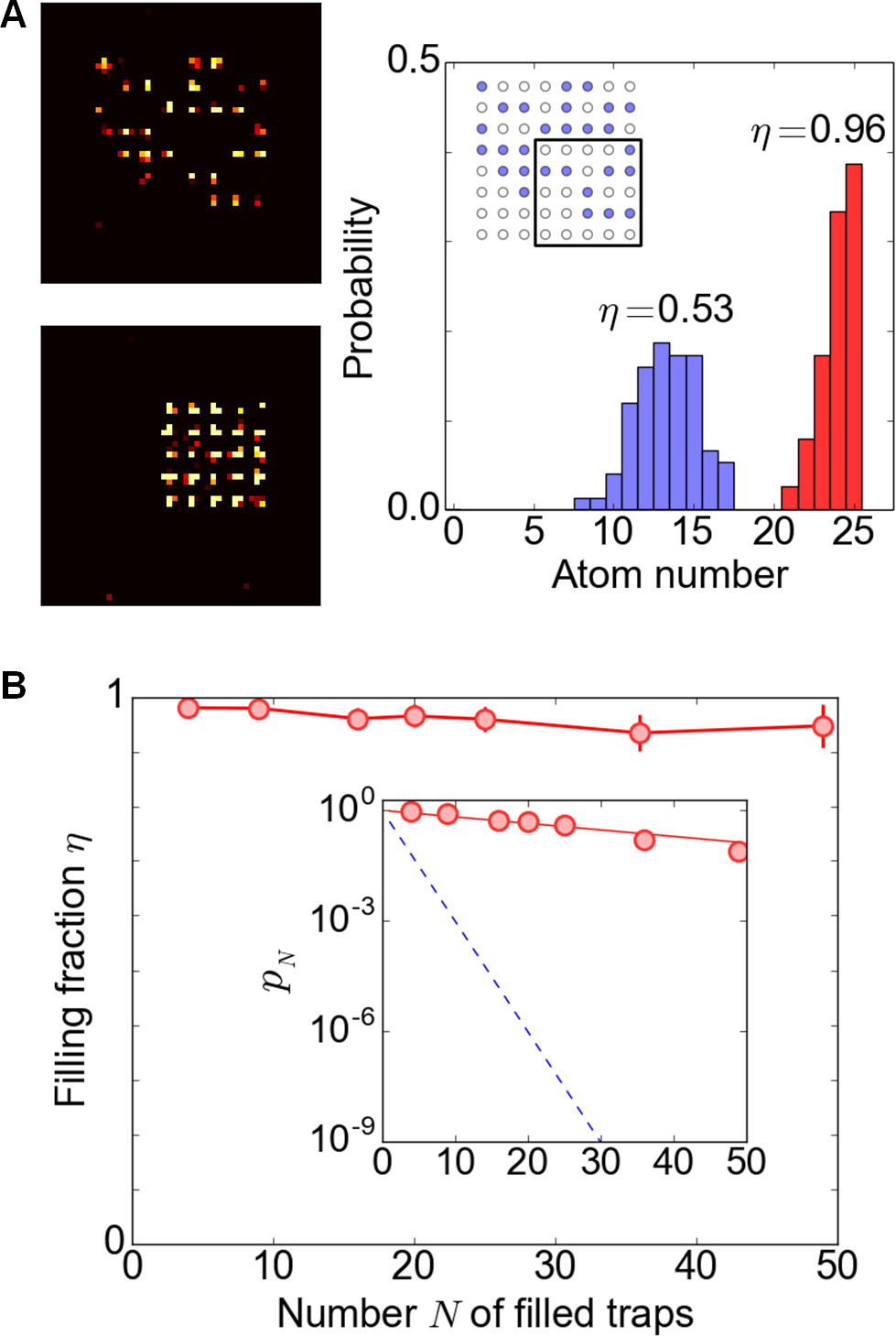}
\caption{{\bf Loading statistics of reordered arrays.} {\bf (A)}: Distribution of the number of atoms in the target $5\times5$ square array  (black square in the inset) before (blue) and after sorting (red), in 100 repetitions of the experiment. The sorted array has $\eta = 96$\% filling fraction (with a standard deviation of 4\%), and defect-free arrays are obtained in $\sim40\%$ of the cases. {\bf B}: Evolution of the filling fraction $\eta$ as a function of the number of traps in the sorted array. The error bars are the standard error on the mean. Inset: observed probability $p_N$ to obtain a defect-free array of $N$ atoms (circles; the dashed line shows the result $p_N=2^{-N}$ corresponding to random loading). }
\label{fig:fig3}
\end{figure}

To implement the atom-sorting shown schematically in Fig.~\ref{fig:fig1}(C), we have used, depending on the nearest-neighbor distance $a$ in the target arrays, two types of elementary moves to transfer an atom from a source trap to a target trap~\cite{SOM}. ``Type 1'' moves transfer an atom by moving it in between other traps, which allows us to have on the order of $N/2$ atom moves only. However, type 1 moves were found to lead to atom loss in the traps close to the moving tweezers trajectory when $a<5\;\mu{\rm m}$, and thus, for such arrays, we used ``type 2'' moves where atoms are moved along the links of the arrays. 

Choosing the best algorithm to calculate the atom moves would ideally require finding the optimal list that minimizes the number of moves and total distance (thus minimizing the time it takes to reorder the array): in our arrays trapped atoms have a vacuum-limited lifetime $\tau_{\rm vac}\sim 10$~s, and thus, the estimated lifetime of a configuration with $N$ atoms is $\tau_{\rm vac}/N$. The total sequence should remain shorter than this to achieve high fidelities. However, finding the optimum set of paths is a hard computational task, reminiscent of the notoriously difficult traveling salesman problem. We have developed a heuristic path-finding algorithm which first computes a list of all possibles moves, and orders it according to their length; we then select moves from source to target traps, starting by the shortest ones, until all target traps are filled. This results in $\sim N/2$ moves (however, this does not necessarily minimize the total travel distance). We finally get rid of the unused extra atoms (if any) by moving them to positions far away from any trap. Additionally, for ``type 2'' moves, we have to avoid moving the tweezers over filled traps (to prevent the tweezers from dragging along other atoms during its motion). In this case, we enforce this constraint in the following way. If the source atom $S$ needs to be moved to the target trap $T$ (we denote this move by $S\to T$) and that an ``obstacle'' $O$ is on the way, we replace the move $S\to T$ by the two moves $O\to T$ and $S\to O$. This increases slightly the number of needed moves, as sometimes some filled traps are ``in the way'', but the overhead is moderate~\cite{SOM}.

For our largest arrays, the total arrangement process is typically performed within less than 50 ms after the initial image is acquired, a timescale still shorter than the lifetime of the initial configuration. We want to emphasize that despite being quite simple and non-optimal, our algorithm is efficient and versatile. Finding better algorithms might however be important for scaling up our approach to hundreds of atoms. 

Figure~\ref{fig:fig2} shows a gallery of trap arrays with arbitrary, user-defined geometries relevant for quantum simulation, e.g., one-dimensional chains, ladders, lattices with square, triangular, honeycomb or kagome structures \cite{Labuhn2016}. Neighboring traps are separated by distances $3<a<6 \,\mu$m. In (A), type-1 moves were used, while in (B), type-2 moves were used. For each array, we show on the top panel a fluorescence image of single atoms obtained with the CCD camera at the beginning of the sequence. Since the probability for each trap to be filled is $p \sim 0.5$, the arrays are initially half-filled. In the accompanying bottom image  we show the final fluorescence image after the sorting is completed. Analyzing 100 repetitions of the experiment for a $5\times 5$ square target array [Fig.~\ref{fig:fig3}(A)], we achieve a filling fraction  $\eta > 96$\%, which gives rise to a probability of getting a defect-free array of about 40\%. As shown in Fig.~\ref{fig:fig3}(B), the filling fraction decreases only marginally when the number of atoms increases, showing the scalability of our approach. In order to achieve even higher filling fractions, one could envision to iterate the procedure presented here, i.e. skip the disposal of unused atoms, analyze the ``final'' image and fill in defects (if any) with remaining atoms. 

In conclusion, we have demonstrated the implementation of a robust procedure based on site-selective atom manipulation, that allows for the rapid preparation of defect-free arrays of single neutral atoms.   Analyzing the technical limitations of the current implementation suggests that preparing hundreds of individual atoms in arrays of arbitrary geometries very close to unit filling is realistic with state-of-the-art technology \cite{SOM}. These results, possibly combined with Raman sideband cooling of atoms in optical microtraps \cite{Kaufman2012,Thomson2013}, open promising paths to study many-body physics and constitute an important resource for quantum information processing with cold neutral atoms. In the future, using the same technique, it should be possible to insert atoms one at a time into a microtrap~\cite{Miroshnychenko2006b}, thus preparing small samples with an exact atom number, e.g. for applications in cold chemistry.

\clearpage
\newpage

\onecolumngrid
\begin{center}
\large{\bf Supplementary material}
\end{center}
\vskip1cm
\twocolumngrid

% For section headers starting with S
\renewcommand{\thesection}{S.\arabic{section}}
\renewcommand{\thesubsection}{\thesection.\arabic{subsection}}
\renewcommand{\thefigure}{S.\arabic{figure}}
\renewcommand{\thetable}{S.\arabic{table}}

\setcounter{figure}{0}

% Hack for making SOM Equations Conform to Science Format
%
% e.g. (S1), (S2), etc
% Requires AMS
\makeatletter %% With ams
\def\tagform@#1{\maketag@@@{(S\ignorespaces#1\unskip\@@italiccorr)}}
\makeatother

% Hack for making figures Say \figurename S\thefigure, e.g. Figure S1:
\makeatletter
\makeatletter \renewcommand{\fnum@figure}
{\figurename~\thefigure}
\makeatother

\makeatletter
\makeatletter \renewcommand{\fnum@table}
{\tablename~\thetable}
\makeatother

\section{S.1 Experimental details}

\begin{figure}[b!]
\centering
\includegraphics[width=8.5cm]{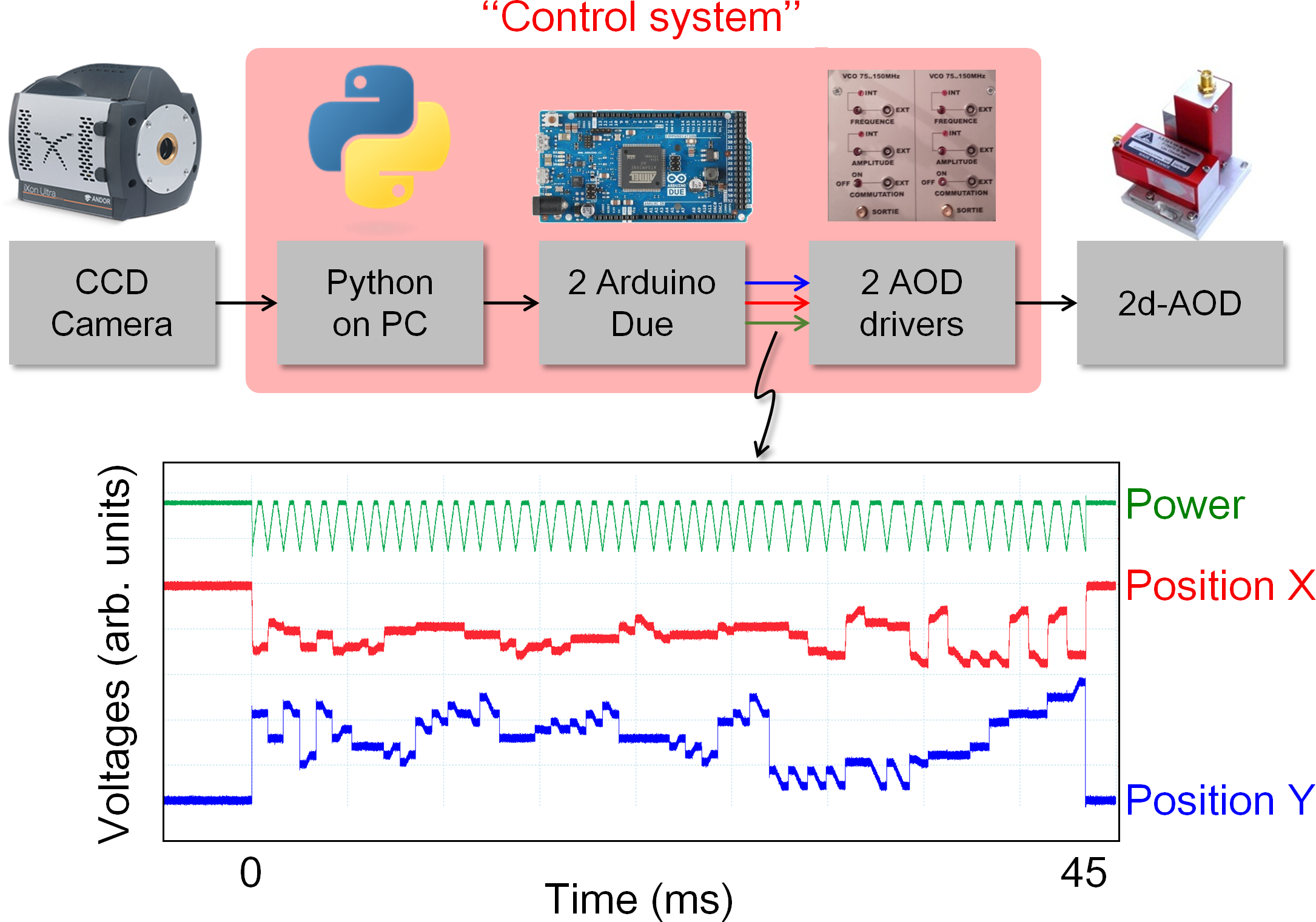}
\caption{{\bf Implementation of the atom assembler.} From the initial image acquired by the camera, a Python program runs the algorithm described in the text and sends to the Arduino Due microcontrollers the coordinates of the source and target traps for all moves. These in turn generate three waveforms (see inset) that, when fed into home-built AOD drivers, control the intensity and X and Y positions of the moving tweezers. The inset shows an example of the waveforms delivered by the microcontrollers for a series of 47 moves.}
\label{fig:figs1}
\end{figure}

Figure S.1 shows a schematic description of the elements of the atom assembler, detailing in particular the ``control system'' depicted as a red box in Figure~1 of main text. Fluorescence images (exposure time 50~ms) of the atoms are acquired by a sensitive CCD camera (we actually use an EMCCD camera [Andor iXon Ultra] but electron-multiplication is not enabled). From these images, a Python program locates the initially filled traps, and, using the algorithm described in the main text, generates a list of moves (i.e. of the pairs of initial and final traps) needed to sort the array. For our largest arrays this algorithm takes less than 1 ms on a standard computer (Intel(R) Core(TM) i5 3470 3.2 GHz, 8 GB RAM; the algorithm is implemented in Python 3.4). The program then transfers this list via the serial port to two microcontrollers (Arduino Due, AT91SAM3X8E ARM Cortex-M3 Board) which convert this list into three waveforms (using 12-bit DAC outputs) that are used to control the intensity and positions along X and Y of the moving optical tweezers.  using home-built RF drivers that feed a two-dimensional AOD (AA-Optoelectonic DTSXY-400-850). The moving tweezers covers a range of $\sim 180 \times 180 \,\mu\rm{m}^2$ in the plane where the atoms are trapped and can be deflected at speeds higher than 1 $\mu\rm{m} /\mu\rm{s}$, much faster than required for adiabatic atom transport. An accurate calibration of the voltage/position conversion is obtained by imaging the plane of the atomic array on a diagnostic CCD camera and measuring the position of the moving tweezers with respect to the fixed traps when the X- and Y- voltages are varied; this ensures that all atom extractions and releases are made at optimal positions. 

\begin{figure}[b!]
\centering
\includegraphics[width=8.5cm]{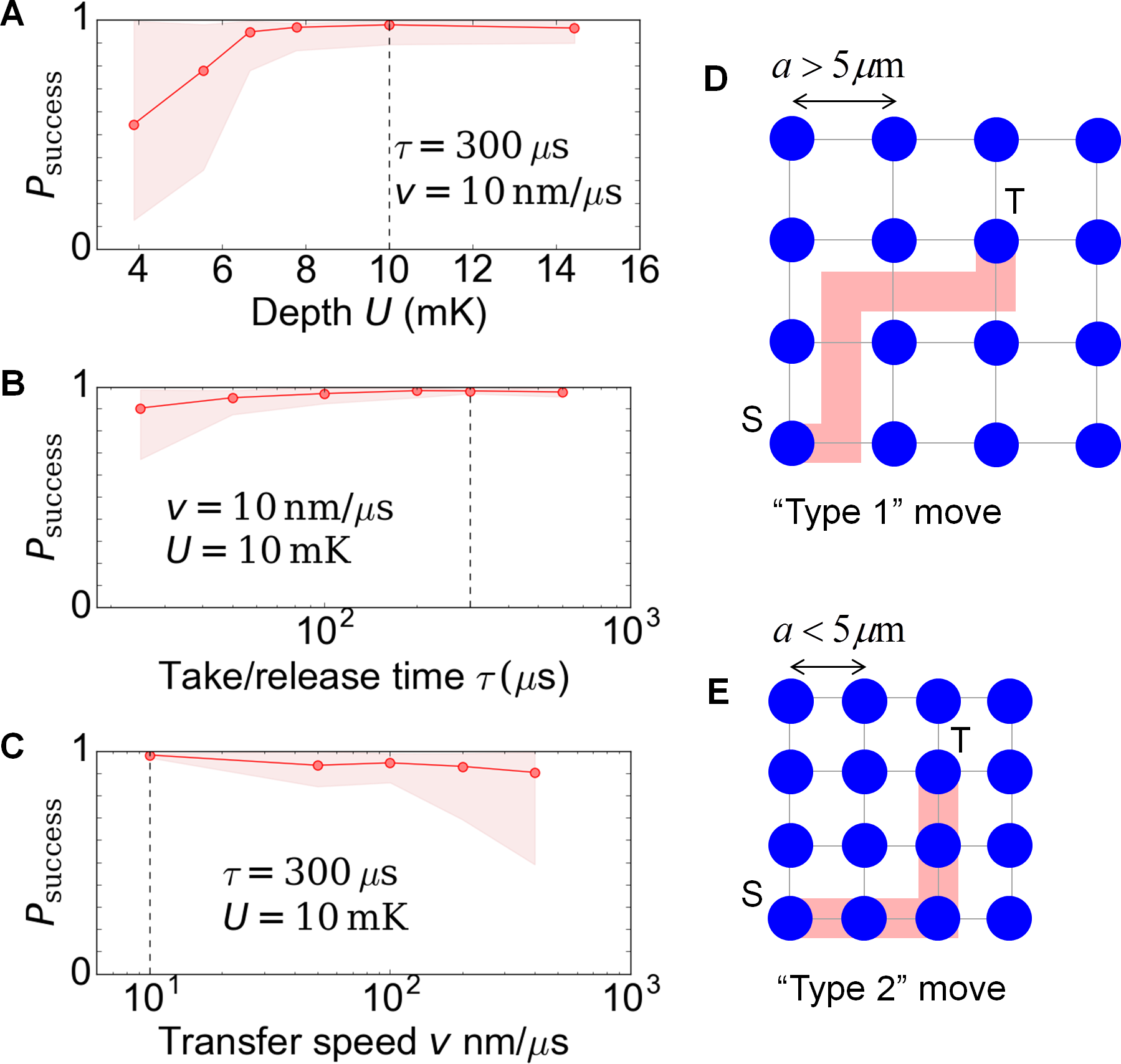}
\caption{{\bf Moving an atom from a ``source'' trap S to a ``target'' trap T.} {\bf (A,B,C)}: Success probability $P_{\rm success}$ for a transfer between two traps separated by $4\;\mu{\rm m}$. {\bf (A)}: as a function of the depth $U$ of the moving tweezers; {\bf (B)}: as a function of the take/release time $\tau$; {\bf (C)}: as a function of the transfer velocity $v$. The points show the success probability averaged over 9 pairs of traps in a square array, the shaded area displaying the full dispersion over the array. The dashed lines indicate the values of $U,\tau,v$ used in the rest of this study. {\bf (D,E)}: Basic moves implemented in the atom-sorting algorithms. {\bf (D)}: if the nearest-neighbor distance in the array is large enough (typically $a>5\,\mu{\rm m}$), we move directly the tweezers from the source to the target, passing in between adjacent atom rows (``Type 1'' moves). {\bf (E)}: otherwise the atoms are moved along the lattice links (``Type 2'' moves).}
\label{fig:figs2}
\end{figure}

\section{S.2 Transfer of an atom between two traps}

We have investigated how the transfer efficiency between two adjacent traps depends on the moving tweezers depth $U$, on the take/release time $\tau$, and on the transfer velocity $v$ [Fig. S.2 A,B,C]. The points are the average of the success probability over different paths in a square lattice  of 9 traps (the shaded area shows the full extent of the distribution over the array). For $\tau=300\;\mu{\rm s}$, a moving tweezers depth of around $U\sim 10U_0=10$~mK,  and a transfer speed of $v\sim 10\;\rm{nm}/\mu\rm{s}$, we measure, in 2000 repetitions of the experiment, a success rate of $0.993(1)$. With these parameters we do not observe appreciable heating during transport (the measured temperatures before and after a move are $T\simeq 50\;\mu{\rm K}$). We use transfer times considerably longer than required for adiabatic motion, but still short enough to allow us to complete a full set of operations to assemble arrays containing tens of atoms. Further optimization could ultimately push time constants down to the limit given by the oscillation frequencies of the trapping potential (using an arbitrary waveform generator to directly drive  the AODs with optimized sweeps, we have measured similar transport fidelities for transfer times ten times as short; however, no attempt was made to incorporate these optimized sweeps into our protocol). 

The two types of moves between a source and a target trap, that we use depending on the nearest-neighbor distance $a$ between two traps, are illustrated in Fig.~S.2. D,E. We observe that both types of move have the same transfer efficiency.

\section{S.3 Prospects for scalability}

\begin{figure}[t!]
\centering
\includegraphics[width=6cm]{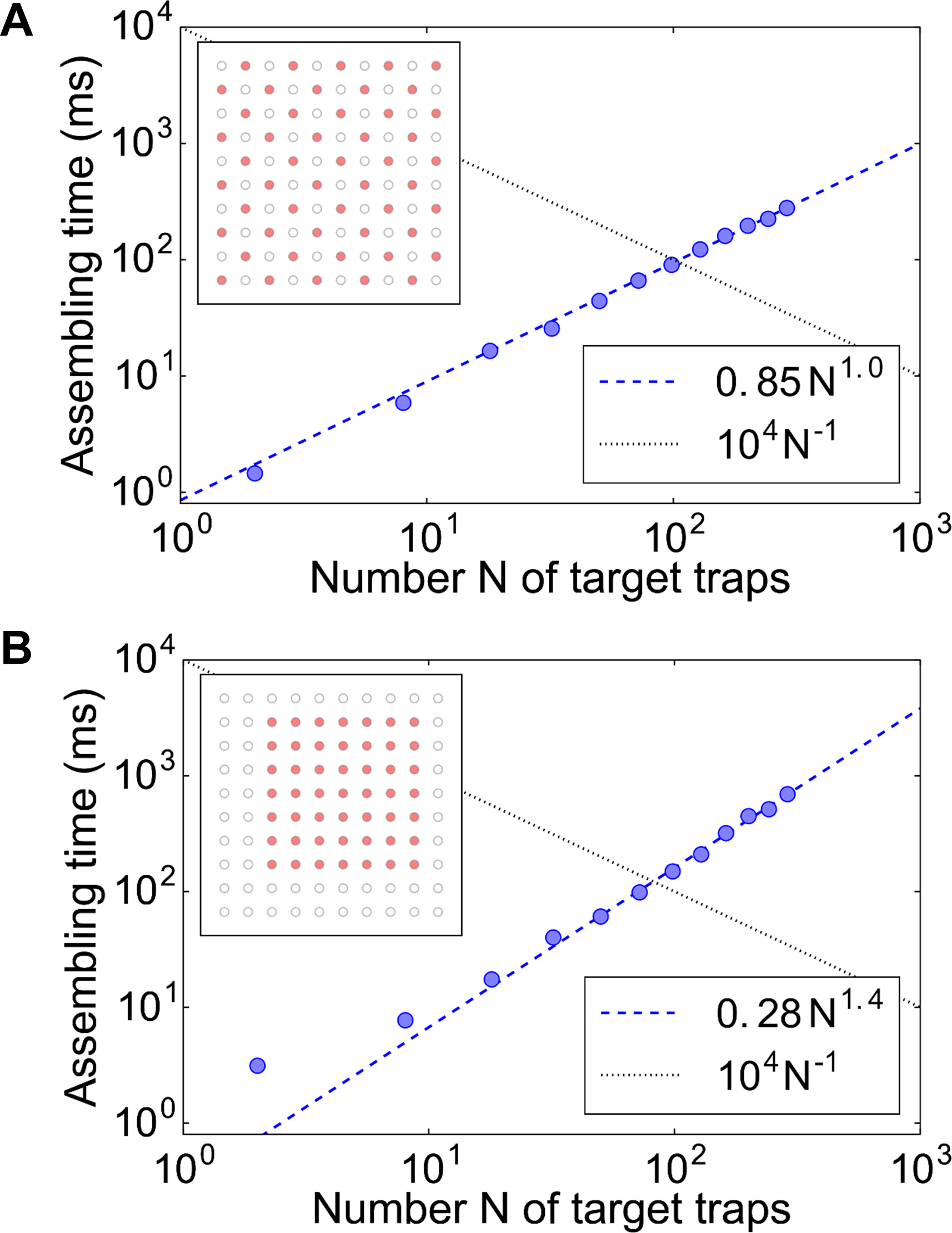}
\caption{{\bf Simulated assembly time versus the number $N$ of atoms in the target array, using ``type-2'' moves.}  {\bf (A)}: For a square lattice in this ``checkerboard'' pattern, the assembly time (points) scales linearly with $N$ (dashed line) for large $N$. {\bf (B)}: For a compact square lattice, the assembly time scales approximately as $N^{1.4}$ (dashed line). This can be understood qualitatively as follows: in such a compact configuration, a significant fraction $\propto N$ of the atoms needs to be moved several times (about $\sqrt{N}$ times, i.e. the linear dimension of the array) between its initial and final position, as other atoms in the way impede a direct transfer. The black dotted lines show the lifetime of the final configuration, assuming $\tau_{\rm vac}=10$~s.  }
\label{fig:figs3}
\end{figure}

Finally, we briefly discuss the scalability of our approach towards a larger number $N$ of atoms.
\begin{itemize}
\item An array of $2N$ traps requires about $10 N$ milliwatts of laser power, meaning that arrays of several hundreds of atoms should be achievable with commercially available lasers delivering a few watts.
\item With our current aspheric lens, coma (and other aberrations) degrade the quality of the tweezers that are more than $\sim20\;\mu$m off-axis, limiting the number of traps to about 200. This could be extended significantly, e.g. by using a microscope objective instead of a single aspheric lens.
\item Another limit is imposed by collisions with the background gas. A defect-free configuration of $N$ atoms has a lifetime $\tau_{\rm vac}/N$, where $\tau_{\rm vac}$ is vacuum-limited lifetime of a trapped atom ($\tau_{\rm vac}\sim 10$~s in our setup). The time needed to assemble the arrays needs to be shorter than this lifetime. Figure S.3 shows the results of simulations (using the experimental values used throughout this work) of the time needed to assemble square arrays of $N$ atoms, as a function of $N$ (blue points), together with the lifetime $\tau_{\rm vac}/N$ of the corresponding final configuration (black dotted line). The intersection of the two curves gives a rough estimate of the array sizes that can be reached without being limited by background-gas collisions, i.e. about $N\sim100$ for our parameters. This can be increased (i) by improving the vacuum (a factor of ten increase in $\tau_{\rm vac}$ is realistic with standard techniques); (ii) by optimizing the duration of each move and minimizing the number of moves required for sorting.
\end{itemize}
The above analysis thus shows that with realistic technical improvements, our approach can be extended to arrays containing several hundreds of atoms. 

\section{S.4 Illustration of the algorithm}

A video illustrating the algorithm on an example of array assembly is available \href{https://www.dropbox.com/s/xy6cc8egcx5gi1m/atom_by_atom_assembler.avi?dl=0}{here}.

\end{document}